\documentclass{aastex}

\usepackage{apjfonts}
\usepackage{emulateapj5}

\begin{document}

\title{External Shear in Quadruply Imaged Lens Systems} 

\author{Gilbert P. Holder$^1$ and Paul L. Schechter$^{1,2}$}
\affil{$^1$ Institute for Advanced Study, Einstein Drive, Princeton NJ 08540}
\affil{$^2$ Department of Physics, Massachusetts Institute of Technology,
77 Massachusetts Avenue, Cambridge, MA 02139}

\begin{abstract}
We use publicly available N-body simulations and semi-analytic models
of galaxy formation to estimate the levels of external shear due to 
structure near the lens in gravitational lens systems. We also describe 
two selection effects, specific to four-image systems, that enhance
the probability of observing systems to have higher external shear.
Ignoring additional contributions from ``cosmic shear'' and assuming
that lens galaxies are not significantly flattened, 
we find that the mean shear at the position of a quadruple
lens galaxy is 0.11, the {\em rms} shear is $\sim$0.15,
and there is a $\sim$45\% likelihood of external shear greater than 0.1.  
This is much larger than previous estimates and in good agreement
with typical measured external shear. 
The higher shear primarily stems from the tendency
of early-type galaxies, which are the majority of lenses, to reside in
overdense regions.
\end{abstract}

\section{Introduction}
\label{sec:intro}

With the benefit of hindsight it can be argued that even the second
gravitational lens discovered, PG1115+080 (Weymann et al. 1980),
foreshadowed what has turned into an embarrassment of riches -- a
superabundance of quadruply imaged quasars.  But only with the advent
of systematic lens surveys (King and Browne 1996; Rusin and Tegmark
2001) has it become clear that the high ratio of quadruple to double
systems is not an artifact of observational selection and therefore presents a
genuine challenge to our understanding of lensing of galaxies.
\nocite{weymann80,rusin01,king96} 
Furthermore, individual fits on a system-by-system basis often require
large amplitude (0.1-0.3) external shear 
\citep{hogg94,schechter97,kneib00,fischer98}, significantly higher
than values of 0.02-0.05 expected 
(Keeton, Kochanek and Seljak 1997; henceforth KKS)
from large scale structure or nearby galaxies.  

\nocite{blandford87,kochanek87,turner84}

There are three factors that will probably have some part to play in the 
ultimate resolution of these problems: galaxy ellipticities, shear due
to random superpositions of mass along the line of sight, and shear
due to structures that are associated with the lens galaxy. In this
paper we concentrate on the shear from associated structures. 

The relative importance of tides and ellipticity has been considered
by KKS.  In computing
the expected tidal shear they consider three contributions: a) random
shear due to unassociated foreground and background structure, b) the
effect of associated galaxies, through the two point correlation
function, and c) the effect of clusters of galaxies, with a term
proportional to the number density of clusters. 
\nocite{keeton97}

In the present paper we take a different tack to estimate the expected
tidal shear due to nearby structures.  We use the GIF
Project recipe (Kauffmann et al. 1999) for galaxy formation within a
cold dark matter (CDM) simulation to compute the shear expected along
random lines of sight, in the directions of galaxies, and specifically
in the directions of early-type galaxies, which appear to be the
predominant type of lens galaxy. In this way, 
the effects of correlated structure are naturally included, without
artificial distinctions between different types of structures, and we
can also include the clustering properties of different types of galaxies.
This is similar to recent work by \citet{white01}, where 
a high resolution hydrodynamical simulation was used to find typical
values of external shear at the positions of typical galaxies, rather
than specifically at the positions of massive early-type galaxies.

In \S~2 we outline our methods for using the GIF simulations to estimate
the effects of correlated structure on the statistics of external
shear in gravitational lens systems. In \S~3 we describe two selection effects
which will further increase the typical external shear measured in quadruple
gravitational lens systems; high shear systems have a larger cross-section
for quadruple lensing (which is partly offset by ``magnification bias'') and
regions of high external shear are likely to be regions of higher convergence,
due to large scale structure. We finish with a discussion of the consequences
of our calculations for various problems associated with
quadruply imaged systems.

\section{Large Scale Structure and Typical Shear}
\label{sec:lss}

Studies of cosmic shear have noted {\em rms} shear of a few percent
\citep{bacon01,vanwaerbeke01,wittman00} from studies of weak distortions of
background galaxies. This result can not be directly applied to estimates
of typical shear values for lens systems because a strong gravitational
lens is not at a random position in the sky but is instead at the position of
a fairly massive galaxy, typically an early-type (elliptical or S0) galaxy 
\citep{keeton98}. 
Such galaxies are known to be located preferentially in overdense regions, 
where it would also be expected that the typical shear would be higher.

To estimate the effects of large scale structure that is correlated with
the lens galaxies, we used the publicly available simulations from the
GIF Project\footnote{\tt http://www.MPA-Garching.MPG.DE/GIF}. The simulations provide
the dark matter distribution as well as estimates of the positions,
velocities, luminosities and colors of galaxies \citep{kauffmann99}.
The galaxy information is derived from a semi-analytic model of galaxy
formation. The simulation box was 141.3 $h^{-1}$Mpc (comoving)
on a side, the 256$^3$ particles each had a mass of 
1.4$\times 10^{10} h^{-1} M_\odot$,
the gravitational softening length was $20 h^{-1}$kpc, and the cosmological
model was a flat universe with $\Omega_m=0.3, \sigma_8=0.9,h=0.7,
\Gamma=0.21$. 

We made projected mass maps of the outputs at $z=0.42$
at a resolution of 50 $h^{-1}$\,kpc and used 
these maps to generate shear maps, assuming a source at $z_s=3$. The
critical surface density for lensing for our assumed cosmology is
approximately 2800\,$h\,M_\odot\, {\rm pc}^{-2}$. At this redshift the 
box size corresponds to $\Delta z \sim 0.05$ and our comoving
50 $h^{-1}$\,kpc resolution element corresponds to an approximate angular 
size of $\sim 8''$, much larger than the typical Einstein radius of
a galaxy ($\sim1''$).

The relation between
convergence (mass surface density in units of the critical surface density)
and shear are algebraic in Fourier space (e.g., Kaiser and Squires 1993), 
so we used an FFT method to generate shear maps from convergence maps.
\nocite{kaiser93}
The Fourier transforms of the convergence map, $\kappa$, and the
two shear components, $\gamma_i$, are simply related:
\begin{equation}
\tilde{\gamma}_1(u,v) = {v^2-u^2 \over u^2+v^2}\ \tilde{\kappa}(u,v) \quad ;
\quad 
\tilde{\gamma}_2(u,v) = {-2uv \over u^2+v^2} \ \tilde{\kappa}(u,v) \quad ,
\end{equation}
where a tilde denotes a quantity in the Fourier domain and $u$ and $v$
are Fourier space variables (i.e., wavenumbers). After making the convergence
map from the particle data of the N-body simulation, the map was Fourier
transformed, the algebraic operations were performed to construct the
shear components in the Fourier domain and the Fourier shear maps were
inverse transformed to produce shear image maps.

With the GIF catalog, we measured 
the shear at all positions that were marked as having massive
early-type galaxies. Our criteria for being a massive early-type galaxy
were $B-V>0.9$ and $M_{star}>10^{11} h^{-1}M_\odot$. We also measured
the shear at the position of all galaxies in the GIF catalog, as well
as at a comparable number of random points in the map.

The measured shear at random points, shown in figure 1,
was typically less than 1\%, as
would be expected from a box with redshift extent of approximately
$\Delta z \sim 0.05$. The few percent cosmic shear is built up from
having many volumes of this extent along the line of sight.
Typical shear values around galaxies were noticeably higher, while
the mean shear at the position of bright early-type galaxies was yet

\centerline{{\vbox{\epsfxsize=8cm\epsfbox{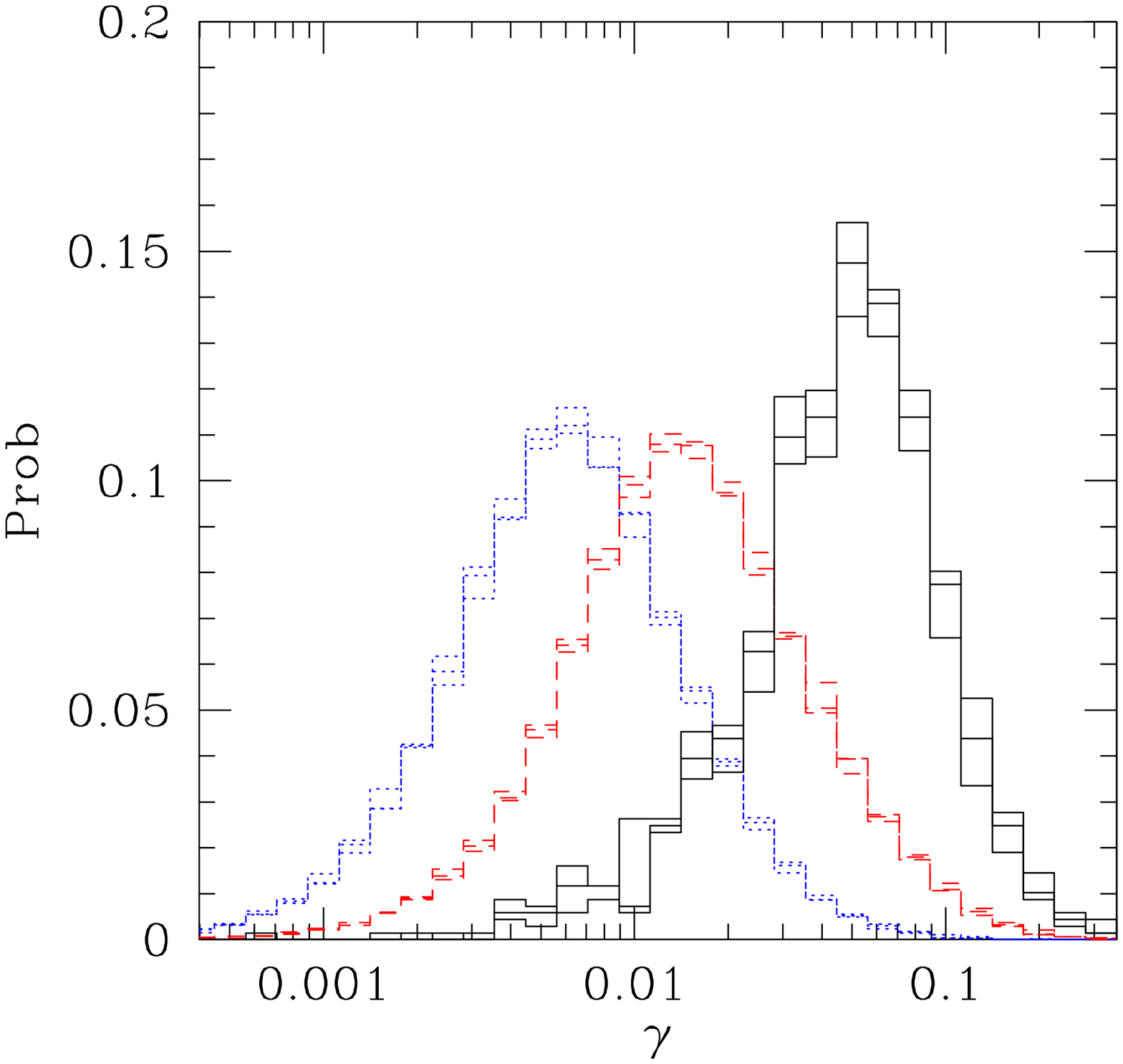}}}}
\figcaption{
Effects of galaxy bias on typical shear in lens systems. 
Black solid curves show shear distributions at position of
early-type galaxies, red dashed curves show shear distributions at position
of all galaxies, blue dotted curves show shear distributions 
at random positions in image. 
Three curves for each type are for three projections of the
simulation volume (along the simulation's x,y and z axes). 
\label{fig:gal}
}
\vskip 0.1in

\noindent higher, 0.058, with an rms shear of 0.071
and an extensive tail past $\gamma \sim 0.1$.  The probability of shear
greater than 0.1 is 0.16.  This is a consequence
of early-type galaxies being preferentially found in or near clusters
of galaxies. Previous estimates (e.g., KKS) took into account that
galaxies were correlated with other galaxies in estimating external
tidal effects, but ignored the preferential location of lenses inside
larger structures in estimating the contribution from large
scale structure. This is an important consideration for two reasons.
More massive galaxies, which have a higher cross-section for producing 
gravitational lens systems, are strongly correlated with large scale structure.
Furthermore, early-type galaxies are even more highly
correlated with large scale structure than late-type galaxies. 

Thus we see that the material correlated with the lens galaxies can
contribute a significant amount of shear. 
The results presented
here are probably an underestimate of the likelihood of high shear events,
as the finite box size results in missing long-wavelength power. 
Other material along the line
of sight will be much less correlated, but will add a significant
amount of {\em rms} shear (KKS find typically $\sim 3\%$ for this
cosmological model).  
The typical shear at the position of a massive early-type galaxy is therefore 
likely larger than $\sim 7\%$.  Note that this is applicable to all 
gravitational lens systems (and early-type galaxies in general) and that we 
have not introduced any distinctions between two-image and four-image systems.

In making the maps smoothing was done on a scale of approximately 
50 $h^{-1}$\,kpc,
reducing the effects of nearby galaxies. There is a trend in
our results for increasing shear with increasing resolution, suggesting a
non-negligible amount of shear from near neighbors, as also suggested
by \citet{white01}; with a smoothing
length of 200 $h^{-1}$\,kpc the mean shear at the position of early-type
galaxies drops to 0.03. The larger smoothing primarily reduces the probability
of high shear values, perhaps suggesting that higher resolution would
lead to an even larger probability of high shear.  In the case of the
gravitational lens system B1422+231 \citep{hogg94} it appears that most of
the shear ($\ga 0.2$ in this case) does in fact come from neighbors 
closer than 50 $h^{-1}$\,kpc.

When all particles within a spherical radius of 50 $h^{-1}$\,kpc  
(comoving) from the centers of 
early-type galaxies were removed before creating the initial mass map
the statistics were nearly unchanged; 
for example the mean shear decreased by only 0.003. 
When this radius of excision was doubled the mean shear was reduced to
0.045, a fractional reduction of more than 20\%. This agrees with both the
above discussion on smoothing and the work of \citet{white01}, and points to
the importance of substructure that is associated with the lens galaxy but 
in the outer regions of the dark matter halo.

\section{Further Selection Effects}
\label{sec:select}

The previous section demonstrated that not all positions on the sky are equally
likely to host a bright early-type galaxy; it is also true that not 
all positions on the sky hosting bright early-type galaxies are equally
likely to host a quadruple gravitational lens system. 
The cross-section for quadruple
lensing by a galaxy can be enhanced by external shear, while the magnification 
statistics are affected by both external shear and convergence from large scale
structure. A bright early-type galaxy in a region of high shear and high
convergence has a higher probability of being observed to be
a quadruple gravitational lens.

In this section, we calculate the relative number of quadruple images
expected in samples, taking into account three effects: 
1) cross-section for quadruple lensing increases with increasing shear; 
2) magnification for quadruple systems decreases with increasing shear;
3) magnification increases with increasing convergence.

The cross-section for producing a quadruple lens is simply the solid angle 
subtended by the part of the source plane that is able to produce four 
observable images (and a fifth highly demagnified, 
thus very difficult to observe, 
image) for a given lens position; i.e., the cross-section is the area 
inside the ``diamond caustic.''  
For a singular isothermal sphere (SIS) lens model in the presence of
external shear, the cross-section for producing a quadruply imaged system 
scales as $\gamma^2/(1-\gamma^2)$\citep{finch02}. 
The probability of
an object producing four images therefore scales roughly as $\gamma^2$,
increasing the probability of galaxies in the presence of high
external shear being in a catalog of observed lenses.  

The magnification is important through the effects of ``magnification bias''
\citep{turner84}. The number of quasars increases toward
lower unmagnified flux; high magnification increases the possible number of 
observable sources. The effects of magnification bias are often in the
opposite direction of the change in cross-section.  Calculation 
of the expected number of lens systems requires the number of sources 
per unit sky area (in the source plane) that could be sufficiently magnified 
to enter the sample.

The number of sources per {\em source plane} solid angle 
expected above a given observed flux limit $S_{min}$, 
in a region of magnification $\mu$, and assuming a power-law 
unmagnified flux distribution
$dN/dS = N_\circ (S/S_\circ)^{-\alpha}$ is 
\begin{equation}
N (> S_{min}) = \int_{S_{min}/\mu}^{\infty} N_\circ 
\Bigl({S \over S_\circ}\Bigr)^{-\alpha} 
dS \quad .
\end{equation}
Employing a change of variable, this reduces to 
simply $\mu^{\alpha-1}$ times the number per source plane
solid angle expected in the
absence of magnification. For a single lens system, there
is a distribution of magnifications in the source plane, so
the factor by which magnification increases
the number of observable quasars is then $<\mu^{\alpha-1}>$, where the
average is over the source-plane distribution of magnifications. 
For  the special case of $\alpha=2$, close to the observed 
value \citep{rusin01}, the factor by which the number of observable
sources per source plane solid angle is increased 
(often called ``magnification bias'') is equal to the mean magnification.

The mean magnification of a quadruple system
for an SIS model in the presence of external shear \citep{finch02} 
scales as $\gamma^{-1}(1-\gamma^2)^{-1}$.  Magnification bias,
in the presence of external shear, therefore has the effect of reducing 
the number of observed lens systems, but by a factor that is
smaller than that by which the cross-section is increased. The net effect of
external shear is to enhance the number of quadruple lenses, 
and we would expect instances of high external shear to be 
overrepresented in catalogs of quadruple gravitational lenses 
relative to massive early-type galaxies in general or two-image lenses.

The effect of additional convergence $\kappa_s$ is to scale the magnifications
by a constant factor $(1-\kappa_s)^{-2}$ and to enhance the effects of
external shear, causing an ``effective shear'' \citep{paczynski86}
$\gamma_{eff}\equiv \gamma/(1-\kappa_s)$. In practice, for the shear and
convergences due to large scale structure the difference between 
$\gamma$ and $\gamma_{eff}$ is negligible.

If we approximate the effects of magnification bias as scaling the fraction 
of lenses by the mean magnification (i.e., assume $\alpha=2$), the net
effect (including cross-sections and magnification) of external shear and 
additional convergence is 
that the probability of a system being observed (assuming a SIS lens with
external shear) to be a quadruple lens 
is proportional to $\gamma_{eff}(1-\gamma_{eff}^2)^{-2} (1-\kappa_s)^{-2}$. 
The tendency of early-type galaxies to reside in overdense regions, where
both external shear and additional convergence are likely,
leads to a significant selection effect in the direction of
enhancing the typical shear at the position of a quadruple image lens system.

This is shown in figure 2, where we show the effects of this selection on
the shear (true shear, not effective shear)
expected for a typical four-image lens system, assuming that the
lens can be modeled as an SIS with external shear. As the input
shear distribution we used the results from the previous section, using
only the contribution from the material in the simulation volume and not
including the contribution from other material along the line of sight. 
The shear distribution was then constructed, with each galaxy position
weighted by the appropriate quadruple lens probability.
The mean convergence at each galaxy position was measured using the original
mass map smoothed on a scale of 300 $h^{-1}$\,kpc to ensure that the
additional convergence was truly from large scale surrounding material.
This will underestimate the extra convergence expected at the position
of lens galaxies, but significantly reduces shot noise as well as 
ambiguity concerning what should be considered part of the lens galaxy.

Including these selection effects, the mean shear expected from only 
the immediately surrounding material is 0.11 and the rms shear is
0.15, with a $\sim$45\% probability of $\gamma>0.1$. 
Therefore, it is to be expected that measured shear in typical
four-image systems should be $\ga 0.1$.

\vskip 0.1in
\centerline{{\vbox{\epsfxsize=8cm\epsfbox{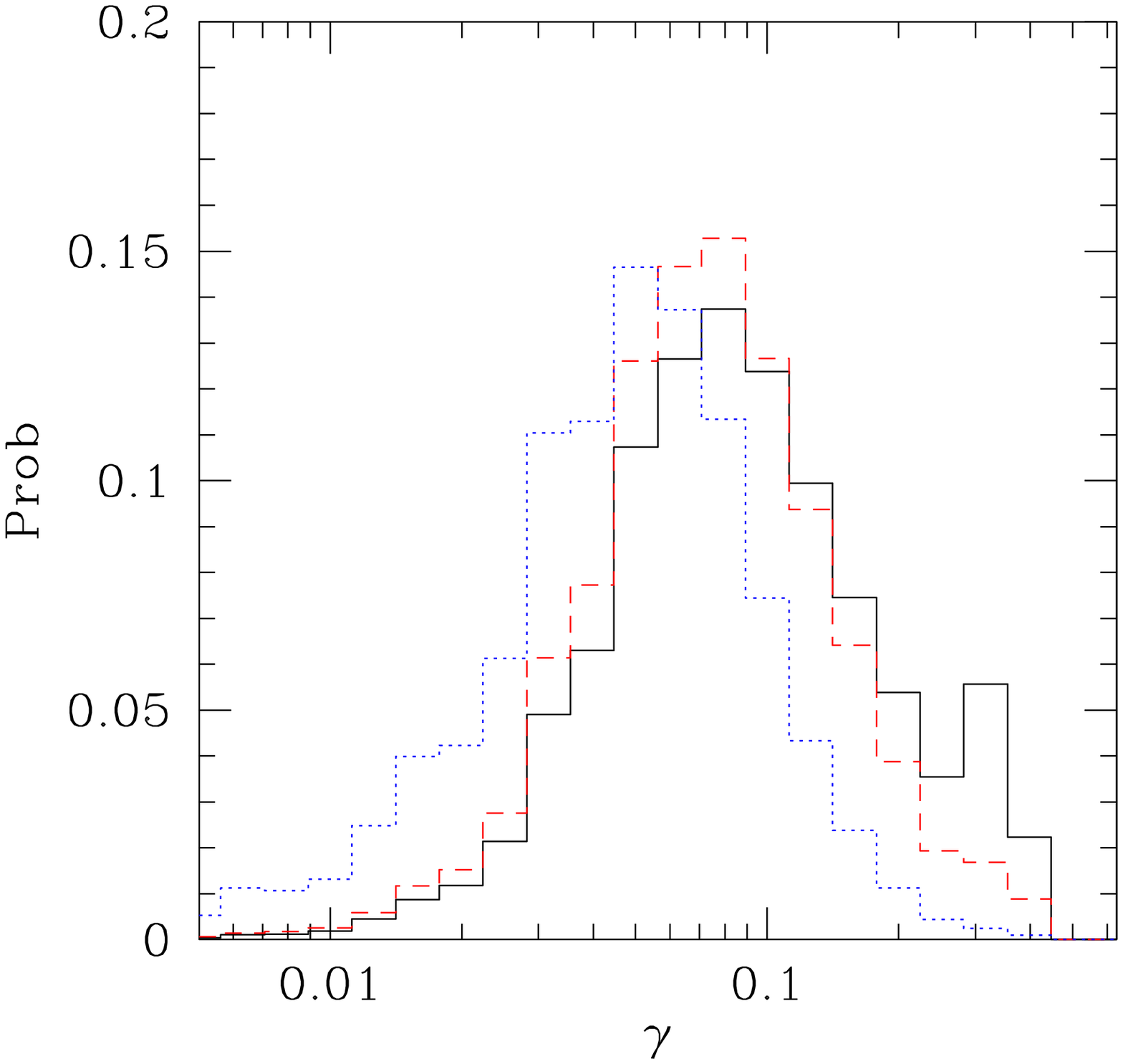}}}}
\figcaption{
Effects of observational biases on observed average shear in
four-image lens systems, assuming a lens model with a SIS and external
shear. Blue dotted curve is shear distribution
at position of early-type galaxies (same as the solid curves in figure 1), 
red dashed curve shows selection effects due to external shear only, 
black solid curve shows combined effects of external shear and extra 
convergence.  Each curve shows the average histogram of the three viewing
projections.
\label{fig:obs}
\vskip 0.4in
}

We have assumed that galaxy ellipticity is not important, but 
ellipticity of the lens galaxy can significantly reduce the enhancement 
of the cross-section due to external shear. For galaxies more elliptical
than an axis ratio of about 0.7:1 the cross-section is nearly unaffected by 
external shear (Rusin and Tegmark 2001). Observed samples of
early-type galaxies in nearby clusters \cite{jorgensen95} contain many
significantly flattened objects, so if typical lens galaxies are
morphologically similar to typical early-type galaxies (of all masses) 
in galaxy clusters, then the selection effects outlined in this section
will be reduced.

\section{Outstanding Issues in Quadruply Imaged Systems}

Previous studies have concluded that external shear is not a viable solution to
the apparent excess of quadruples (Rusin and Tegmark 2001 and references
therein) but on a system-by-system basis it is 
apparent that external shear is significantly higher in lens systems than had 
been predicted. The calculations in the previous sections demonstrate that the
external shear expected in a typical quadruple lens system is roughly 0.10, 
with a large tail extending to $\sim 0.3$, 
much larger than previously predicted. 
The effects of large scale structure uncorrelated with the lens, 
structures within 50 $h^{-1}$\,kpc of the lens, and substructures 
within the lens galaxy will increase typical shear values further.

Using the compilation of lens models from the CASTLES web 
site\footnote{\tt http://www-cfa.harvard.edu/glensdata},
the mean reported
external shear (for all lens systems, not just quadruple systems) is 0.10, 
with an rms of 0.12. This should be treated
as anecdotal evidence for typical shear of $\sim$0.1; several models
have a non-positive number of degrees of freedom, the reported
uncertainties have been ignored in our calculations (i.e., not inverse variance
weighted) and many of the fits have been performed holding the
galaxy ellipticity fixed. We have shown that such a distribution of
shears can comfortably be accommodated within the current understanding
of galaxies and their environments. The observation that lens galaxies 
are predominantly early-type galaxies is fully consistent with the 
observed external shears of $\sim 0.1$. As would be expected, the observed 
shear in quadruple-image systems appears to be systematically larger than
for double-image lenses.

To solve the excess of four-image systems compared to two-image systems,
typical external shear on the order of 0.3 is 
required (KKS; Finch {\em et al.} 2002).
We are in agreement with previous work that such high shears are not expected 
from currently viable cosmological models. 
A combination of external shear and galaxy
ellipticity is not efficient for increasing the fraction of quads,
as they are most likely not correlated and therefore do not add
coherently.  The main effect of external shear in the presence of 
a distribution of ellipticities is to increase the fraction of 
quadruple images only in the low ellipticity regime. 

Other possible solutions to the problem of too many quadruple images can
be enhanced by the addition of external shear. For example, adding a
core to the mass profile or making the central density profile less 
steep can enhance the fraction of quadruple images \citep{rusin01}. 
The additional convergence due to large scale structure will introduce a 
moderate selection effect, as the image separations will be increased, making
it more likely that small splittings could be resolved.
However, it does not appear that external shear can by itself solve the 
problem of the apparent excess of four-image systems.

\section{Discussion}
\label{sec:disc}

We have shown that the
relatively high amounts of external shear found in most lens systems 
naturally arise from non-linear 
large scale structure in the vicinity of the lenses.
Early-type galaxies are systematically in more overdense regions and therefore
would be expected to experience higher amplitude tidal fields.

The fact that quadruple lens galaxies are mainly early-type galaxies 
is enough to explain the observed values of external shear, 
previously considered high.  The dominance of early-type galaxies
in lens samples arises simply from the bias in lens samples toward more
massive galaxies, which are more likely to be early-type galaxies
\citep{fukugita91,maoz93,kochanek93}.

This amount of shear expected from nearby structures is not sufficient to, 
by itself, solve the ``quad problem'' (the apparent relative 
super-abundance of four-image systems) but the effect goes in the right
direction and should add with other effects, such as making density
profiles in the centers of galaxies less steep.
Paying careful attention to the preferred positions of galaxies relative
to typical places in the universe, as we have done here, will almost
certainly have an important role in understanding these observations.

Typical shear values are sufficiently high that external shear should be
of comparable or greater importance in lens modeling to internal shear
(galaxy ellipticity), in agreement with model fits to known lenses. 
We do not expect the introduction of galaxy ellipticities to significantly 
change our conclusions, but it would be a natural direction for future work.

Selection effects are important in gravitational lens systems. The results
presented here suggest that our knowledge of structure formation and
galaxy environments can be usefully applied to understanding the observed
statistics and characteristics of lens systems.  As numerical simulations
of large scale structure, models of galaxy formation, 
and gravitational lens catalogs all continue to improve we can expect the 
importance of an integrated approach to only increase.

\acknowledgements{GPH is supported by the W.M.
Keck Foundation. PLS is grateful to the John Simon Guggenheim
Foundation for the award of a fellowship. PLS is supported by
grant NSF AST 02-06010 and was supported by NASA HST-GO-06555.01-A.  
We are very grateful to the Virgo
Consortium and the GIF project for making their materials
publicly available. The simulations in this paper were carried out 
by the Virgo Supercomputing Consortium using computers based at 
Computing Centre of the Max-Planck Society in Garching and at the 
Edinburgh Parallel Computing Centre. The data are publicly 
available at www.mpa-garching.mpg.de/NumCos. We thank Scott Gaudi for 
many useful discussions.
}

\end{document}